# Performance comparison of multi-user detectors for the downlink of a broadband MC-CDMA system

F. Portier***, R. Legouable*, L. Maret **, F. Bauer ****, N.Neda *****,
J-F. Hélard***, E. Hemming****, M. des Noes**, M. hélard*
*France Télécom R&D: 4 rue du clos courtel, 35510 Cesson-Sévigné, France
rodolphe.legouable/maryline.hélard@francetelecom.com
**CEA-LETI: 17 rue des Martyrs 38054 Grenoble cedex 09
mathieu.desnoes/luc.maret@ cea.fr
***IETR/INSA: 20 Ave. des Buttes de Coësmes, 35043 Rennes Cedex, France
fabrice.portier/ jean-francois.helard@insa-rennes.fr
**** Nokia Research Center, Meesmannstr. 103, 44807 Bochum, Germany
franziskus.bauer/erwin.hemming@nokia.com
***** Centre for Comm. Syst. Research, University of Surrey,Guildford, Surrey  GU2 7XH
n.neda@surrey.ac.uk

**ABSTRACT**

In this paper multi-user detection techniques, such as Parallel and Serial Interference Cancellations (PIC & SIC), General Minimum Mean Square Error (GMMSE) and polynomial MMSE, for the downlink of a broadband Multi-Carrier Code Division Multiple Access (MC-CDMA) system are investigated. The Bit Error Rate (BER) and Frame Error Rate (FER) results are evaluated, and compared with single-user detection (MMSEC, EGC) approaches, as well. The performance evaluation takes into account the system load, channel coding and modulation schemes.

## I. INTRODUCTION

Since 1993, a modulation technique called MC-CDMA has been proposed for multimedia services in high data rate wireless networks [1]. This promising multiple access scheme with high bandwidth efficiency combines the CDMA as a multiple access technique and the OFDM as a Multi-Carrier transmission system. Unfortunately, it suffers from multiple access interference (MAI) which limits its performance.

In this paper, the performance of various multi-user detectors that mitigates MAI for the downlink of a MC-CDMA transmission are evaluated.

In section II, the system parameters defined in the IST MATRICE project [2] are described.

In section III, the multi-user detectors (MUD) that have been evaluated are described, and in section IV the simulation results are given. Eventually, section V summarises the results and draws conclusions.

## II. SYSTEM DESCRIPTION

The general scheme of the MultiCarrier (MC) system used in this work is depicted in Figure 1. The convolutional or turbo encoded, punctured and interleaved data signal is modulated with either QPSK or 16 QAM and fed into a discrete Hadamard Transform for spreading. The resulting chips are then frequency interleaved and distributed over the whole bandwidth. This signal is then processed into a N points IFFT, provided with a guard interval and transmitted into the channel. At the receiver the guard interval is removed and the signal is transformed into the frequency domain by an FFT. In the investigations presented herein perfect channel knowledge is assumed. The frequency domain signal and the channel coefficients are handed over to the multi-user detector modules that are compared in this contribution. The output of the MUD module is despread and soft demapped. The soft bits are deinterleaved, depunctured and then processed by a channel decoding unit, which is either a convolutional/Viterbi or a turbo decoder.

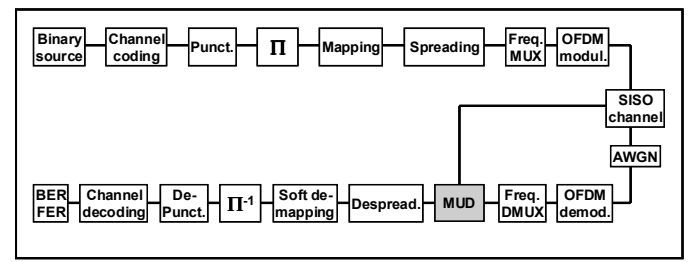

**Figure 1: System Overview.**

Representing the transmission process in the frequency domain, the MC system states in a very simple form:

$$\mathbf{y} = \mathbf{H}\sum_{k=0}^{K-1}\mathbf{x}_k + \mathbf{n} \qquad (1)$$

where $\mathbf{x}_k$ is the $N{\times}1$ vector of the signal transmitted by user k, $\mathbf{y}$ is the $N{\times}1$ vectors representing the received signal, $\mathbf{n}$ is the additive white Gaussian noise vector with $E\{\mathbf{n}\mathbf{n}^H\} = \sigma_n^2 \cdot \mathbf{I}_N$ where $\mathbf{I}_N$ is the identity matrix, and $\mathbf{H}$ is a diagonal matrix collecting the channel frequency response at subcarriers frequencies :

$$\mathbf{H} = diag\{H_0 \quad H_1 \quad \cdots \quad H_{N-1}\} \text{ with } H_k = H\!\left(e^{j\frac{2\pi k}{N}}\right)$$

A general expression for a downlink MC-CDMA system is the following:

$$\mathbf{x} = \sum_{k=0}^{K-1} \mathbf{x}_k = \left(\mathbf{I}_{N_u} \otimes \mathbf{C}\right)\mathbf{d} \quad (2)$$

where $\mathbf{C} = (\mathbf{c}_1, \ldots, \mathbf{c}_K)$ is $S_F \times K$ the matrice containing all the used spreading codes and $\mathbf{d}$ is the ($KN_u \times 1$) vector collecting the transmitted symbols ($S_F$ is the spreading factor, $K$ the number of active codes, $N_c$ is the number of modulated carriers and $N_u = N_c / S_F$). They are grouped by $N_u$ blocks of $K$ symbols:

$$\mathbf{d} = (\mathbf{d}_1, \mathbf{d}_2, \ldots, \mathbf{d}_{Nu})^T$$

$\mathbf{d}_m = (d_{1m}, d_{2m}, \ldots, d_{Km})^T$ is the ($K \times 1$) vector collecting the K symbols to be transmitted in $m^{th}$ sub-band. Expression (2) is depicted schematically in Figure 2.

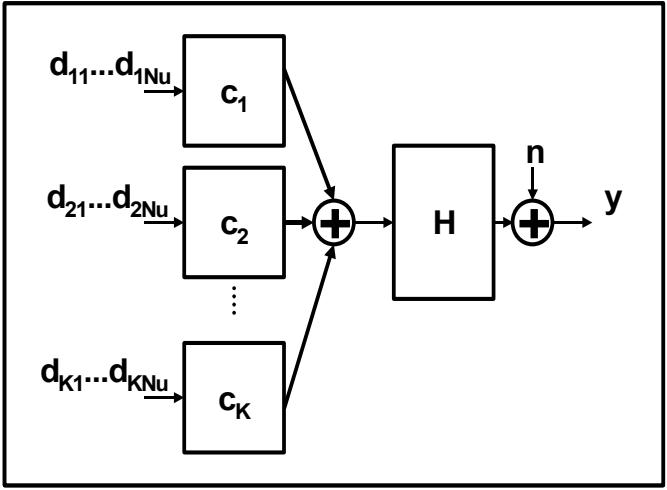

**Figure 2: Frequency transformed, schematic view of MC-CDMA.**

In the sequel, the equalizer works on each subband separately. Thus the received signal in the $m^{th}$ subband is:

$$\mathbf{y}_m = \mathbf{H}_m \mathbf{C} \mathbf{d}_m + \mathbf{n} \quad (3)$$

where $\mathbf{H}_m = \text{diag}(H_{mNu}, \ldots, H_{mNu+SF-1})$. For shake of simplicity, the index m will be dropped.

The received signal $\mathbf{y}$ includes the code multiplexed symbols of all users. As the channel disrupts the orthogonality of the codes multiple access interference (MAI) occurs, leading to bad decisions at the receiver side.

### III. MUD DESCRIPTION

The multi-user detectors can be split in two families: linear equalizers and interference cancellation schemes.
In this paper the MMSE Combining (MMSEC), Global MMSE (GMMSE) equalizers and a polynomial implementation of GMMSE are evaluated. Then, non-linear Interference Cancellation algorithms are introduced, using Successive or Parallel SUD schemes combined with the despreading process.

*A. MMSEC*

The best performance among Single-User (SU) detectors are obtained with MMSEC, which is a trade-off between MAI reduction (restoring the orthogonality among users), and noise enhancement. It minimizes the mean square value of the error $\|\hat{\mathbf{x}}_k - \mathbf{x}_k\|^2$ and needs information about the channel and SNR over each sub-carrier.

At the receiver side, the equalizer is characterized by a diagonal matrix G followed by a despreading operation:

$$\hat{d}_k = \mathbf{c}_k^H \mathbf{G} \mathbf{y}$$

Equalizer coefficients, applied independently on each carrier, are equal to:

$$g_l = \frac{H_l^*}{|H_l|^2 + 1/\gamma_c} \rho \quad (4)$$

$$\rho = \frac{S_F}{\sum_{n=0}^{S_F-1} \frac{|H_n|^2}{|H_n|^2 + 1/\gamma_c}} \quad (5)$$

where $\gamma_c$ is the signal to noise ratio per sub-carrier, and $\rho$ is the normalization factor in order to cope with high order modulations, as 16 QAM for example.

*B. GMMSE*

Performing the Mean Square Error criterion on the received signal, we obtain a generalized detection joining equalization and despreading [6]. Contrary to MMSEC which works at chip level (per sub-carrier), GMMSE inverts the channel at symbol level, taking into account both the spreading codes and the propagation channel. As well, it makes a trade-off between MAI reduction and noise enhancement.

To minimize $\|\hat{d}_k - d_k\|^2$, the optimal weighting vector, according to Wiener filtering, is:

$$\mathbf{G} = \mathbf{\Gamma}_{yy}^{-1} \mathbf{\Gamma}_{yd_k} \quad (6)$$

where $\mathbf{\Gamma}_{yy}$ is the autocorrelation matrix of the received vector $\mathbf{y}$ and $\mathbf{\Gamma}_{ydk}$ is the cross-correlation between the desired symbol $d_k$ and the received vector $\mathbf{y}$. The sub-carrier noises have the same variance and are independent. In the downlink, since the user signals have the same power ($E\{d_k^2\} = E_s$) and are independent, we can write $E\{\mathbf{dd}^H\} = E_s \cdot \mathbf{I}_K$. Then, the equalization coefficient matrix, assuming a normalized code matrix $C$, is:

$$\mathbf{G} = \mathbf{c}_k^H \mathbf{H}^H \left( \mathbf{H} \mathbf{C} \mathbf{C}^H \mathbf{H}^H + \frac{\sigma_N^2}{E_S} \mathbf{I} \right)^{-1} \quad (7)$$

In case of full load ($K = S_F$), $\mathbf{C} \cdot \mathbf{C}^T = \mathbf{I}_K$, Eq.(7) leads to the same equalizer as MMSEC. On the other hand, when the capacity is not full ($K < S_F$), the equalization coefficient matrix $\mathbf{G}$ is no more diagonal. In that case, the Global MMSE (GMMSE) algorithm outperforms MMSEC, since it minimizes the decision error taking into account the despreading process instead of minimizing the error independently on each sub-carrier. However, whatever the number of active users *K*, this solution implies to solve a $S_F \times S_F$ linear system.

An alternative formulation, which is strictly equivalent, consists in applying the matched filter to received vector *y* before MMSE filtering. With this solution, used for implementation, we get a new expression :

$$\mathbf{G} = \rho \mathbf{e}_k^H \left( (\mathbf{HC})^H (\mathbf{HC}) + \frac{\sigma_N^2}{E_S} \mathbf{I} \right)^{-1} \mathbf{C}^H \mathbf{H}^H \quad (8)$$

where $\mathbf{e}_k$ is the column vector with zeros everywhere excepted in the $k^{th}$ position.

This equalization is normalized in order to cope with high order modulations. The normalization coefficient is:

$$\rho = diag \left( \mathbf{C}^H \left( \mathbf{HC}(\mathbf{HC})^H + \frac{\sigma_n^2}{E_s} \mathbf{I} \right)^{-1} \mathbf{HC} \right) \quad (9)$$

However, by contrast with equation (7), applying equation (8) only implies to solve a $K \times K$ linear system.

*C. polynomial GMMSE*

The optimal GMMSE receiver offers very good performance, but its complexity is high due to the matrix inversion operation. Moreover, if a long scrambling code is used in addition to the Walsh-Hadamard channelization codes, the equalizer has to be computed for every new MC-CDMA symbol. This prevents to use this latter receiver at the mobile terminal. One solution is to replace matrix inversion by a polynomial expansion of this matrix. For practical reasons, this sum shall be truncated, leading to lower performance than the infinite polynomial MMSE receiver:

$$\left( \mathbf{I} + \mathbf{HCC}^H \mathbf{H}^H \right)^{-1} = \sum_{i=0}^{I-1} (-1)^i \left( \mathbf{HC} \mathbf{C}^H \mathbf{H}^H \right)^i$$

The principle of this receiver is to compute the coefficients $(a_k(i))_{i=0,\ldots,I-1}$ of the polynomial which minimize the mean square error $\left\| \hat{d}_k - d_k \right\|^2$ :

$$\hat{d}_k = \mathbf{c}_k^H \mathbf{H}^H \sum_{i=0}^{I-1} a_k(i) (\mathbf{HCC}^H \mathbf{H}^H)^i \mathbf{y}$$

The solution of this minimization problem was first found by Moshavi and al [7]. Unfortunately, the polynomial coefficients depends on a complex way of the spreading codes and channel coefficients which makes its implementation very complex. In this study we implemented a solution developed in [8] for DS-CDMA systems. It consists in applying results from the random matrix theory in order to eliminate the dependence of the polynomial coefficients on the actual code values. Only the property that the codes are orthogonal is taken into account.

*D. PIC / SIC*

In order to handle a large number of users, receivers can also implement sub-optimal non-linear interference cancellation (IC). The principle of IC is to detect the information of the interfering users and to reconstruct the interfering contribution in order to subtract it from the received signal. IC can be performed in parallel for all interfering users with Parallel Interference Cancellation (PIC) detectors, or successively with Successive Interference Cancellation (SIC) detectors where only the strongest interferer remaining after the previous IC stage is cancelled [5]. In our simulations, the SUD technique used at each stage is MMSE, and the number of stages is fixed.

*Successive Interference Cancellation*

The SIC detector first detects the most powerful interfering user and then cancels its contribution from the received signal. The second strongest interferer is then cancelled and so on. The processing may be repeated for a few or for all users. A complete detector would consider all users, but commonly only the interferers stronger than the useful one are suppressed. SIC detector is generally used when the power of some users are higher than the power of the useful user. Since processing one supplementary stage leads to an additive time delay, a trade-off between the number of stages and the total acceptable delay has to be found. The process is carried out iteratively until the remained interferers are considered insignificant. The resulting signal is finally despread. The data detection may be either hard or soft.

*Parallel Interference Cancellation*

The Parallel Interference Cancellation (PIC) structure is based on an estimation of the total interference due to the simultaneous other users in order to remove it from the received signal. The contribution of all interfering users is cancelled in parallel reducing the time delay of a SIC detector. The expression of this iterative system for the $i^{th}$ stage and the $k^{th}$ user is given by the following:

$$\hat{d}_k^{(i)} = \mathbf{c}_k^H \mathbf{G}^{(i)} \left( \mathbf{y} - \mathbf{H} \sum_{\substack{j=0 \\ j \neq k}}^{K-1} \hat{d}_j^{(i-1)}(i) \mathbf{c}_j \right) \quad (10)$$

with the expression of the initial stage given by:

$$\hat{d}_k^{(i)} = \mathbf{c}_k^H \mathbf{G}^{(0)} \mathbf{y}$$

The received signal is first equalised by a SU technique, then it is despread by each code. An Inverse Fast Hadamard Transform (IFHT) can be implemented since the system is synchronous. As for SIC detector, data detection may be either hard or soft. After detection, the data is spread again, tapped by the channel coefficients **H** and then subtracted from the received signal. Finally, the resulting signal with lower MAI term is then equalised, despread and detected. We can note that the second equalizer structure ($\mathbf{G}^{(i)}$) may be different from the first one ($\mathbf{G}^{(i-1)}$).

## III. SIMULATIONS RESULTS

In order to compare all the proposed detection techniques, simulation results have been carried out in terms of Bit error Rate (BER) and Frame Error Rate (FER) according to the system load. The BRAN E propagation channel model, representative of urban environment and defined into the ETSI-BRAN project for HIPERLAN2 [3], has been considered. The simulations have been launched with the following simulation parameters:

- Sampling frequency = 57.6MHz;
- Velocity of 60km/h;
- FFT size N = 1024;
- Guard interval of 216 samples, allowing the absorption of all echoes;
- Number of available modulated carriers $N_c$ = 736, leading to a signal bandwidth of 41.46MHz;
- Walsh Hadamard spreading codes of length of 32 chips are used, leading to $N_u$ = 23 spread data symbols per OFDM symbol;
- The frame structure is composed of 30 OFDM symbols
- Random time and regular frequency interleaving are implemented

Simulation results have been obtained either with the UMTS convolutional or turbo channel coding schemes (CC and TC) [4].

In order to show the behaviour of all techniques according to the system load, some performance results have been extracted from the BER and FER curves to obtain the necessary energy per bit to noise ratio, $E_b/N_0$, to reach a fixed BER or FER according to the number of active users. Figure 3 and Figure 4 present the required Eb/N0 values to reach a BER equals to $10^{-4}$ when using respectively CC and TC schemes. We note that the performance decreases as the system load increases and the MAI is not totally mitigated by the simulated MUD. In Figure 3, for capacity superior to one half, we also remark that MMSEC outperforms the one-stage PIC and global SIC detectors, showing that IC detectors lead to bad decisions into the iterative process, generating errors. In addition, as some errors have been made into the IC process, it generates bad metric computations at the input of the decoder, leading to error propagations at the decoder level. This result confirms the one obtained in [5]. Even if these results have been obtained with implementation of hard decisions into the IC process, no significant gain would have been recorded with soft decisions. However, even if the GMMSE detector gives the best performance, especially at half load, the IC detectors, the MMSEC and GMMSE receivers are very closed in terms of performance (within 0.5 dB), while EGC and asymptotic polynomials receiver performance are worse. As previously mentioned in mathematical formulations, MMSEC and GMMSE techniques have similar performance results at full load.

In Figure 4, as opposed to the QPSK modulation, the receivers can be clearly ranked. The linear detectors (GMMSE and MMSEC) offers the best performance results. PIC and SIC detectors with turbo-coded scheme do not provide good performance at low signal to noise ratios (which is the case with usual turbo codes) due to bad decisions of the estimation of the interferers into the IC process. GMMSE is the scheme, which offers the best results mainly with non full load systems. Compared to the MMSEC scheme, the gain is of 2dB for 3/4 rate turbo coded 16QAM and half loaded system. This gain is significant but leads to an increase of the system complexity. In Figure 4, performance results of EGC and asymptotic polynomials receiver have not been plotted because of high error floor. In fact, EGC and asymptotic polynomials receivers are too much sensitive to MAI with a 16 QAM constellation. Using a large constellation will increase the global level of MAI, and at the same time, the detector is more sensitive to noise level since the distance between 2 points of the constellation is smaller than with a QPSK modulation. Eventually, EGC and asymptotic polynomials receivers are not suited for the downlink of a high bit rate MC-CDMA system.

Figure 5 and Figure 6 present the required Eb/N0 values to reach a FER equals to $10^{-2}$ when using respectively CC and TC schemes until half load. The general behaviour of these FER curves is identical to the BER one, leading to the same conclusions.

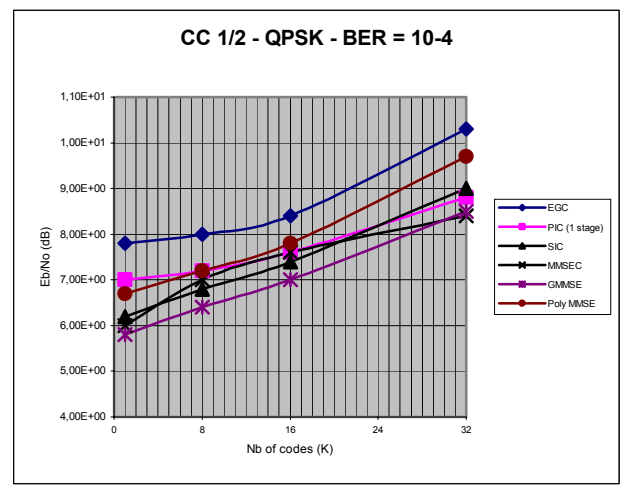

**Figure 3 : Influence of system load on BER for CC and QPSK.**

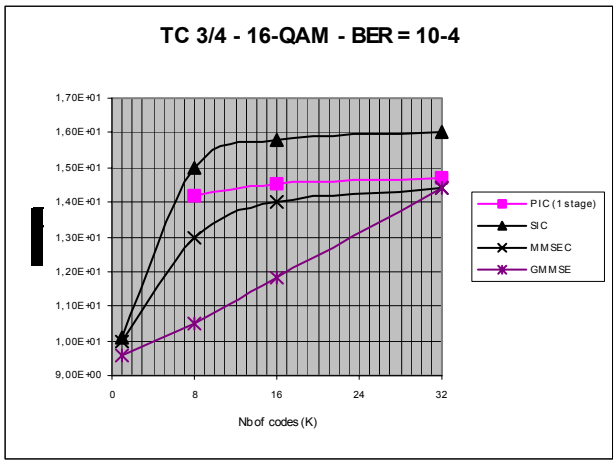

**Figure 4 : Influence of system load on BER for TC and 16-QAM.**

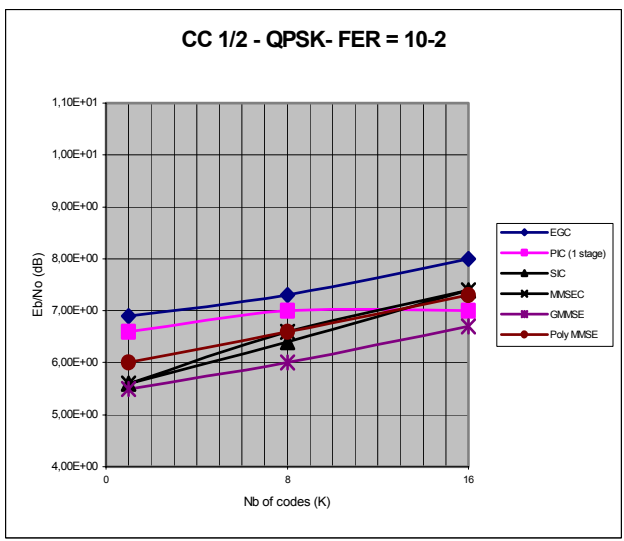

**Figure 5 : Influence of system load on FER for CC and 16-QAM.**

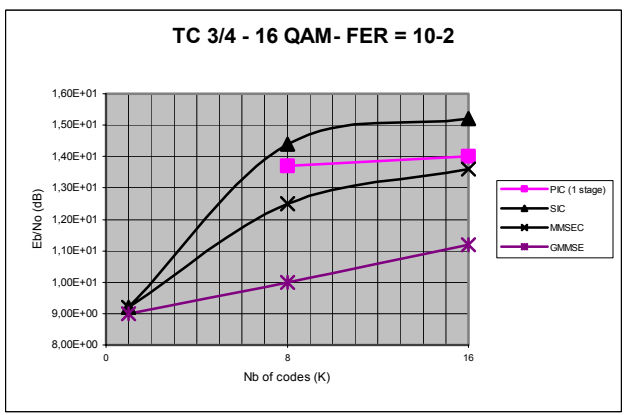

**Figure 6 : Influence of system load on FER for TC and 16-QAM.**

## VI. CONCLUSION

The bit error rate and frame error rate performance of multi-user detection techniques for the downlink of a MC-CDMA system were presented. It was seen that the GMMSE outperforms all other multi-user detection techniques, especially for high bit rate scenarios, whereas the EGC and polynomial MMSE schemes results in very poor performances. However, the GMMSE is computationally excessive. It was also observed that the MMSEC could provide a better trade-off between performance and complexity, especially under high load conditions.

## ACKNOWLEDGEMENTS

The work presented in this paper was carried out in the project Matrice (MC-CDMA Transmission Techniques for Integrated Broadband Cellular Systems) that is supported from the European Commission in the framework of FP5 with the contract number IST-2001-32620. The authors would like to acknowledge for this support and the possibility to carry out the research work.